\documentclass[aps,pra,reprint,groupedaddress]{revtex4-1}

\usepackage{graphicx}
\usepackage{color}
\usepackage{dcolumn}
\usepackage{amsmath}
\usepackage{amssymb}
\usepackage{units}
\usepackage{braket}
\usepackage{natbib}
\usepackage[]{hyperref}
\hypersetup{
    colorlinks=true,
    linkcolor=black,
    citecolor=blue,
    urlcolor = blue
}

\begin{document}

\title{Signatures of spatial inversion asymmetry of an optical lattice observed in matter-wave diffraction}

\author{C. K. Thomas$^{1}$}
\author{T. H. Barter$^{1}$}
\author{T.-H. Leung$^{1}$}
\author{S. Daiss$^{2}$}
\author{D. M. Stamper-Kurn$^{1,3}$}
\thanks{Electronic address: dmsk@berkeley.edu}
\affiliation{
    $^1$Department of Physics, University of California, Berkeley, California 94720, USA\\
    $^2$Max-Planck-Institut f{\"u}r Quantenoptik, Hans-Kopfermann-Strasse 1, 85748 Garching, Germany\\
    $^3$Materials Sciences Division, Lawrence Berkeley National Laboratory, Berkeley, California 94720, USA}

\date{13 June 2016}

\begin{abstract}
The structure of a two-dimensional honeycomb optical lattice potential with small inversion asymmetry is characterized using coherent diffraction of $^{87}$Rb atoms.  We demonstrate that even a small potential asymmetry, with peak-to-peak amplitude of $\leq 2.3\%$ of the overall lattice potential, can lead to pronounced inversion asymmetry in the momentum-space diffraction pattern.  The observed asymmetry is explained quantitatively by considering both Kapitza-Dirac scattering in the Raman-Nath regime, and also either perturbative or full-numerical treatment of the band structure of a periodic potential with a weak inversion-symmetry-breaking term.  Our results have relevance for both the experimental development of coherent atom optics and the proper interpretation of time-of-flight assays of atomic materials in optical lattices.
\

\noindent DOI: \href{http://journals.aps.org/pra/abstract/10.1103/PhysRevA.93.063613}{10.1103/PhysRevA.93.063613} 
\end{abstract}

\maketitle

In x-ray crystallography, the diffraction of light is analyzed to determine the exact crystalline structure of a material.  Similarly, with the availability of ultracold sources of coherent matter waves of atoms, one can use atomic diffraction to characterize potentials experienced by the atoms.  Of particular interest are the optical lattice potentials produced by periodic patterns of light intensity and polarization, formed by the intersection of several coherent plane waves of light or by direct imaging.  Lattice potentials of various geometries and dimensionalities, some incorporating atomic-spin dependence and gauge fields, have been produced or proposed for the purpose of creating synthetic atomic materials by placing quantum-degenerate atoms within them \cite{pets94,bloc05nphys,bloc12qsim}.  Just as in condensed matter, the characteristics of such synthetic atomic materials derive from the nature of the optical crystal upon which they are based.  Matter-wave crystallography therefore becomes a vital tool in the study of such synthetic quantum matter \cite{PortoBandGap}.

A key first step in determining the structure of a lattice is the assignment of its point-group and space-group symmetries.  The violation of a symmetry is identified in x-ray crystallography by a difference in the intensities of diffraction spots \cite{ladd13xraybook}.  Following such work, here we detect the inversion asymmetry of an optical lattice by observing significant asymmetries in the diffraction of a coherent matter wave from the potential.  For this, we produce a spin-polarized $^{87}$Rb Bose-Einstein condensate at rest, and then impose for a variable pulse duration the two-dimensional honeycomb optical lattice potential produced by three light beams intersecting at equal angles \cite{solt11hexagonal}.  The resulting Kapitza-Dirac diffraction is quantified by imaging the gas after it is allowed to expand freely.  By tuning the pulse time and working with a deep optical lattice, we produce highly visible (over $50\%$ contrast) inversion asymmetry in the populations of the first-order diffraction peaks even while the inversion-asymmetric part of the potential is $\leq 2.3\%$ of the overall lattice potential.  This observation highlights the extreme sensitivity of coherent matter-wave scattering in revealing features of a potential landscape under investigation.

Aside from demonstrating sensitive optical-lattice crystallography, our observation also has implications for the development of atom optics.  Matter-wave interferometers for several applications have employed brief pulses of light to split and recombine atomic beams coherently \cite{wicht2002preliminary, kasevich1992measurement}.  Kapitza-Dirac diffraction, i.e., the diffraction of atoms from standing-wave rather than traveling-wave optical potentials, has the advantage that it is technically simple to implement, requiring only light waves at a single optical frequency \cite{mosk83,ovch99}.  However, as compared with Bragg or Raman diffraction, it has the disadvantage of being less efficient and less selective \cite{martin1988bragg}.  The technical simplicity has inspired modifications of Kapitza-Dirac diffraction employing several pulses of light so as to diffract atoms to selected diffraction orders with high efficiency \cite{wu05splitting}, although the diffraction remained inversion symmetric, with as many atoms diffracted to the wave vector $+\mathbf{G}$ as to the wave vector $-\mathbf{G}$.  We show that this last constraint can be lifted to produce inversion-asymmetric Kapitza-Dirac diffraction of matter waves in two dimensions.  Similar to the previous demonstration in one dimension \cite{jo12kag}, we explain how this asymmetry arises from the interference between different diffraction pathways to the same final momentum state.
%
\begin{figure*}[]
\centering
\includegraphics[scale=.8]{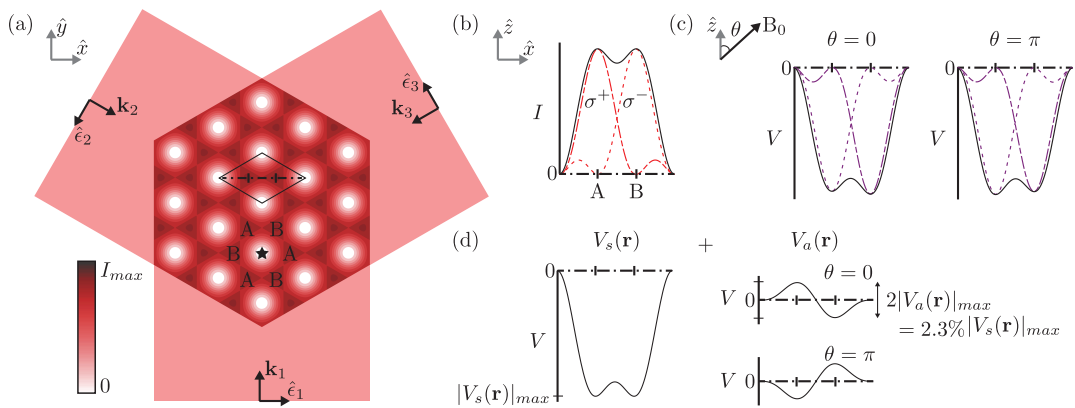}
\caption{Three 1064 nm beams interfere at $120^\circ$ with in-plane polarization to create a honeycomb lattice of intensity maxima.  We identify the unit cell of the lattice potential (solid line).  A dashed line within the unit cell runs through the two potential minima, which are marked with ticks and labeled A and B.  One-dimensional profiles of the light intensity (b) and optical potentials (c) and (d) along this line are shown.  The star symbol, located at a minimum-intensity location, serves as a center for the spatial inversion operation that exchanges the A and B sites of the lattice. (b) We define a quantization axis orthogonal to the lattice plane and show that the light is predominantly $\sigma^+$ at site A and $\sigma^-$ at site B.  (c) The atoms are polarized by a uniform magnetic field B$_0$ at an angle $\theta$ from the quantization axis.  We show the lattice potential for extreme values of $\cos \theta$, where the potential depth at sites A and B maximally differ.  (d) The lattice potential is the sum of an inversion-symmetric potential $V_s(\mathbf{r})$, that arises from the scalar Stark shift and an inversion-antisymmetric potential $V_a(\mathbf{r})$, that comes from the vector Stark shift. }
\label{fig:Scheme}
\end{figure*}

We begin by describing the optical lattice potential characterized in this work.  As in Ref.\ \cite{solt11hexagonal} and illustrated in Fig.\ \ref{fig:Scheme}(a), we form a two-dimensional honeycomb lattice using three beams of light at the wavelength $\lambda = 1064$ nm, with equal intensity, propagating horizontally and intersecting at equal angles, with each beam linearly polarized in the lattice plane.  We define a quantization axis orthogonal to the lattice plane and show in Fig.\ \ref{fig:Scheme}(b) that the beams produce a periodic pattern of varying intensity and optical polarization.

Rubidium atoms exposed to this optical lattice experience an ac-Stark shift that can be divided into scalar, vector, and tensor terms acting on the atomic hyperfine spin \cite{cct72}.  The tensor light shift is negligible in our experiment owing to the large detuning of the lattice light from the atomic transitions.  Figure \ref{fig:Scheme}(d) shows the lattice potentials that result from the scalar and vector parts of the ac-Stark shift.  The scalar light shift is proportional to light intensity and produces a honeycomb lattice potential $V_s(\mathbf{r})$ with two sites of equal depth per unit cell, labeled A and B in the figure.  The vector light shift in the presence of a dominant external magnetic field produces a potential $V_a(\mathbf{r})$ that is approximately diagonal in the Zeeman basis defined by the field direction.  $V_a(\mathbf{r})$ is proportional to both intensity and the dot product of helicity and atomic spin \cite{cct72}.  The helicity in the lattice is staggered so that $V_a(\mathbf{r})$ is of opposite sign at each of the two sites in the unit cell.  

The scalar and vector light shift potentials differ in their inversion symmetry, with $V_s(\mathbf{r})$ being symmetric and $V_a(\mathbf{r})$ being antisymmetric under spatial inversion. Figure \ref{fig:Scheme}(a) shows one of the zero-intensity locations within the optical lattice as an example of the center of the inversion operation.  The  result of this operation is to switch sites A and B.  
 
For alkali atoms, $V_a(\mathbf{r})$ is suppressed with respect to $V_s(\mathbf{r})$ owing to the large optical detuning from the atomic resonance.  For the wavelength of light used in our lattice, the ratio $2 |V_a(\mathbf{r}) / V_s(\mathbf{r})|$ is at most $2.3\%$, so that $V_a(\mathbf{r})$ adds only a small inversion-symmetry-breaking potential atop a graphene-like, inversion-symmetric honeycomb lattice. Within this limit, we control the magnitude and sign of $V_a(\mathbf{r})$ by tilting the dominant external magnetic field $\mathbf{B}_0$ by an angle $\theta$ with respect to the (vertical) axis defined by the optical helicity.  For atoms spin polarized along $\mathbf{B}_0$, the asymmetric potential is then $V_a(\mathbf{r}) \propto \cos\theta$.  Figures \ref{fig:Scheme}(c) and \ref{fig:Scheme}(d) show that the resulting lattice potential has a small, state-dependent offset in energy between sites A and B.

To characterize this lattice using matter waves, we create a nearly pure, optically trapped Bose-Einstein condensate of $3\times 10^{5}$ $^{87}$Rb atoms that is spin polarized in the $|F=1, m_F = -1\rangle$ state along the axis defined by a $\sim$ 0.5 G applied magnetic field.   We then introduce a three-beam lattice potential with $\lvert V_s(\mathbf{r}) \lvert_{max}=h \times 87 \pm 4$ kHz for a pulse time $\tau$ between 10 and 100 $\mu$s.  This lattice depth is calibrated with independent measurements of the diffraction produced by the one-dimensional lattices formed by pairs of the lattice beams \cite{morsch2006dynamics}.    After the pulse, we simultaneously switch off the optical lattice and optical trapping potentials and allow the atoms to expand freely for a 20-ms time of flight.   We finally take an image of the density distribution in which the various diffraction orders, generated at the reciprocal lattice vectors by exposure to the lattice potential, are seen as separate peaks.  

\begin{figure}[t]
\includegraphics[]{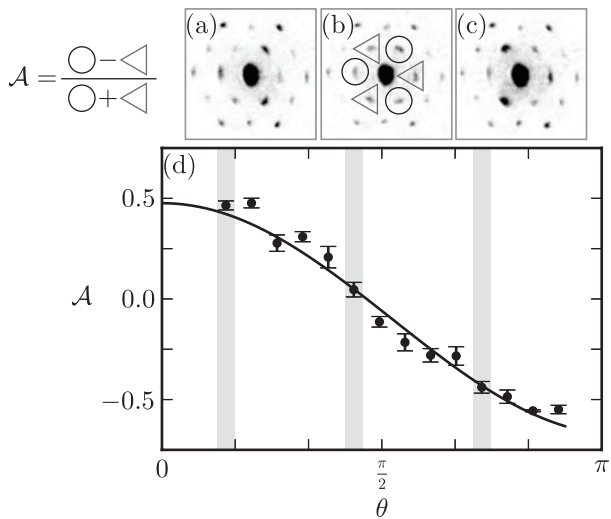}
\caption{An asymmetry parameter $\mathcal{A}$ is defined as the first-order population imbalance and measured for data taken as a function of $\theta$ with a pulse time of $50 \ \mu s$. (a) Time-of-flight image for $\theta = 0.44$ shows an asymmetry in the first-order diffraction peaks.  (b) We highlight the first-order peaks with circles (at $\mathbf{G}_i$)  and triangles (at $-\mathbf{G}_i$).  (c) Time-of-flight image for $\theta = 2.2$ shows reversal of the observed asymmetry.  (d)  $\mathcal{A}$ is computed for each of five images and the mean and standard error of these data are plotted. The solid line shows the expected dependence on $\theta$.}
\label{fig:SpinDep}
\end{figure}

\begin{figure}[t]
\centering
\includegraphics[]{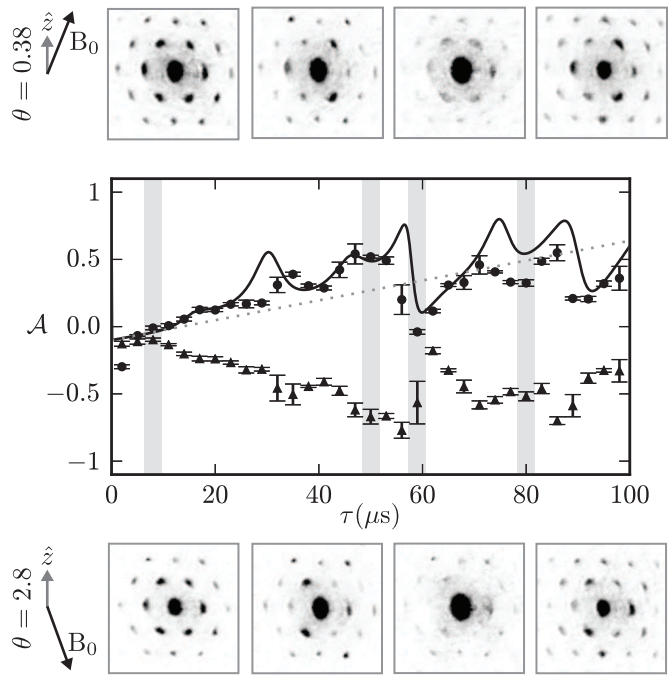}
\caption{Oscillations in $\mathcal{A}$ as a function of the Kapitza-Dirac pulse time $\tau$, shown for $\theta=0.38$ radians (circles) and $\theta=2.8$ radians (triangles).  The data represent the mean and standard error of five experimental runs at each pulse time.  A numerical calculation (solid line) with no free parameters closely reproduces the time dependence of $\mathcal{A}$, while perturbation theory (dashed line) captures the short time behavior. Inset time-of-flight images for $\tau$ of 8, 50, 59, and 77 $\mu$s show directly the evolution of the first-order asymmetry. We note that discrepancies between theory and experiment, e.g., at times around 30, 60, and 80 $\mu s$, appear when the total population in the first-order peaks is small, causing a systematic reduction in the measured magnitude of $\mathcal{A}$.}
\label{fig:TimeDep}
\end{figure}

The first-order diffraction peaks in Figs.\ \ref{fig:SpinDep}(a) and \ref{fig:SpinDep}(c) show a pronounced inversion asymmetry.  To quantify this asymmetry, we identify three reciprocal lattice vectors that describe first-order diffraction as $\mathbf{G}_1 = \mathbf{k}_3 - \mathbf{k}_2$ and its cyclic permutations, where $\mathbf{k}_{1,2,3}$ are the wave vectors of the incident beams that form the lattice.   We define an asymmetry parameter $\mathcal{A}$ as
\begin{equation}
\mathcal{A} = \frac{\sum_i \left(P_{\mathbf{G}_i} - P_{-\mathbf{G}_i}\right)}{\sum_i \left(P_{\mathbf{G}_i} + P_{-\mathbf{G}_i}\right)},
\end{equation}
i.e., as the contrast between the diffraction intensities at wave vectors $\mathbf{G}_i$ and $-\mathbf{G}_i$, the two sets of wave vectors being related by inversion.  This measure is robust against variations in the total atom number and against residual center-of-mass motion of the condensed atoms with respect to the lattice potential.  We note that imaging aberrations introduce a slight offset in $\mathcal{A}$ (of about 0.1) in our experiment, seen in Figs.\ \ref{fig:SpinDep} and \ref{fig:TimeDep}.

We confirm that the momentum-space inversion asymmetry is caused by the real-space inversion asymmetry of the lattice potential by varying the magnitude and sign of the inversion-symmetry-breaking potential $V_a(\mathbf{r})$.  We tune $V_a(\mathbf{r})$ by rotating the orientation of the magnetic field from the vertical axis by the polar angle $\theta$ before exposing the condensate to the lattice potential.

Our data emphasize the fact that even an asymmetry in the lattice potential of $\leq 2.3\%$ can lead to highly visible asymmetry in the matter-wave diffraction pattern.  The evolution of the momentum-space asymmetry $\mathcal{A}$ vs pulse time is portrayed in Fig.\ \ref{fig:TimeDep}.  The asymmetry grows from small values at early times to over $50\%$ at $\tau \sim 50 \, \mu$s, and also displays clear modulation in time reflecting the coherent dynamics of matter waves within the imposed lattice potential.  Throughout these dynamics, reversing the sign of the inversion asymmetry of the lattice reverses the observed inversion asymmetry of the diffracted atoms.


\begin{figure}[!t]
\centering
\includegraphics[]{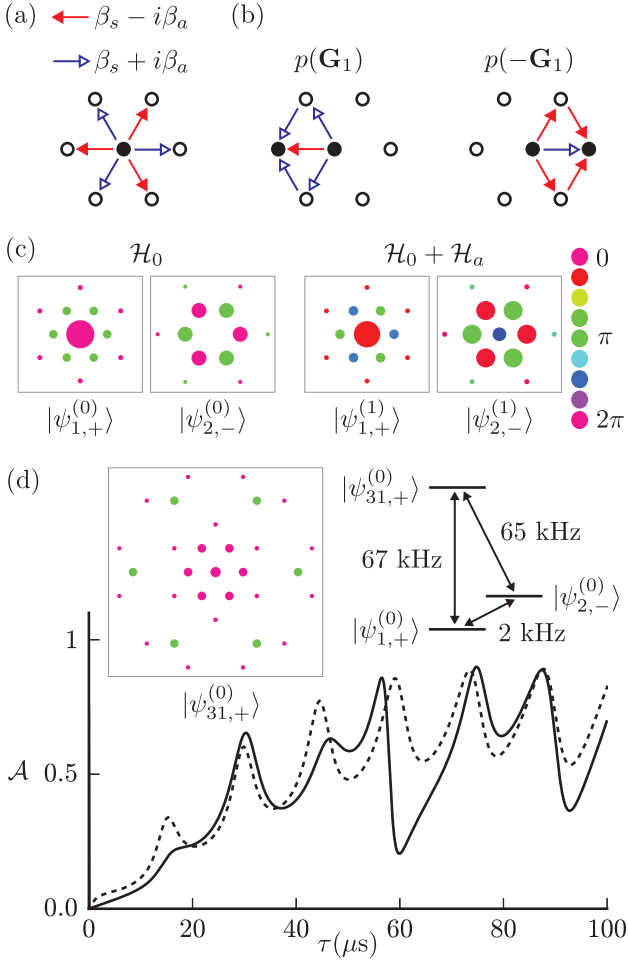}
\caption{(a) Atoms at zero momentum are coupled to wave vectors $\pm \mathbf{G}_i$ by the asymmetric Fourier components of the potential. (b) Interference between first- and second-order processes creates a population imbalance at $\pm \mathbf{G}_i$. (c) We treat the inversion-asymmetric potential as a perturbation $\mathcal{H}_a$ on the inversion-symmetric lattice Hamiltonian $\mathcal{H}_0$ and show momentum-space amplitudes (spot size) and phases (color) of the two lowest energy eigenstates.  In our experiment $\alpha_{2,1}$ is large and  $\mathcal{H}_a$ strongly mixes the symmetric ground state, $|\psi_{1,+}^{(0)}\rangle$, and the antisymmetric excited state,  $|\psi_{2,-}^{(0)}\rangle$.  Both perturbed states are asymmetric and overlap with a stationary condensate.  (d) Much of the oscillatory behavior observed in $\mathcal{A}$ can be attributed to the beating of three states, identified in the eigenspectrum of $\mathcal{H}_0$ as $ \ket{\psi_{1,+}^{(0)}}$, $\ket{\psi_{2,-}^{(0)}}$, and $\ket{\psi_{31,+}^{(0)}} $.  We show state $\ket{\psi_{31,+}^{(0)}} $, which is the excited state that best satisfies the criteria described in the first scenario of the text.  The energy differences among these states define three frequencies (2, 65 and 67 kHz) that dominate the signal of $\mathcal{A}$.  Our numerical calculations show that this three-state description (dotted line) captures most of the physics in the full signal of $\mathcal{A}$ (solid line).} 
\label{fig:ExpFig}
\end{figure}

We present two physical pictures that explain the origin of the observed momentum-space inversion asymmetry.  First, we consider how the momentum-space asymmetry originates from low-order diffraction in the lattice.  This description, shown schematically in Figs.\ \ref{fig:ExpFig}(a) and \ref{fig:ExpFig}(b), is valid in the limit of a shallow optical lattice and in the Raman-Nath regime, where we can ignore the kinetic energy of the diffracting atoms \cite{gadway2009analysis}.  Both the scalar and vector Stark shift optical lattice potentials, $V_s(\mathbf{r})$ and $V_a(\mathbf{r})$, can be characterized in momentum space by their Fourier transforms $V_{s,a}(\pm \mathbf{G}_i)$ at the wave vectors $\pm \mathbf{G}_i$, where the relation $V_{s,a}(\mathbf{G}_i) = V_{s,a}^*(-\mathbf{G}_i)$ is valid because both potentials are real.  Considering the $C_3$ rotational symmetry of both lattices and their respective inversion symmetries we have $V_{s}(\pm \mathbf{G}_i) = \beta_s$ and $V_{a}(\pm \mathbf{G}_i) = \pm i \beta_a$, where $\beta_s$ and $\beta_a$ are both real.

We now consider the probability amplitudes $p(\pm \mathbf{G}_i)$ for atoms diffracting from their initial zero momentum state to a final wave vector $\pm \mathbf{G}_i$ within a time $\tau$.  Figure \ref{fig:ExpFig}(a) illustrates that such diffraction can be achieved by one first-order process, with amplitude $- i (\beta_s \mp i \beta_a) \tau/\hbar$, and by two second-order processes, which sum to an amplitude $(-i)^2 (\beta_s \pm i \beta_a)^2 \tau^2/\hbar^2$.  We ignore higher order terms.  Interference between the first- and second-order scattering amplitudes results in an imbalance of probability for diffraction into opposite wave vectors.  Calculating the asymmetry parameter $\mathcal{A}$ at short times and for small lattice asymmetry ($|\beta_a| \ll |\beta_s|$) we obtain $\mathcal{A} \simeq 6 \beta_a t/\hbar$, which is plotted as a gray dotted line in Fig.\ \ref{fig:TimeDep} and describes the data well for small $\tau$.

While the model above provides a simple analytic expression for $\mathcal{A}$, its assumptions are violated under the conditions of our experiment. For one, our experiments are performed with a deep lattice that leads to diffraction to high order, as exemplified by the many diffraction peaks in our images.  Second, the measurements are performed with pulse times that are long enough to be outside the Raman-Nath regime, which is shown by the high kinetic energy of the large momentum states produced in our experiment.  Therefore, the diffraction pattern produced in our measurement is better described as resulting from coherent dynamics governed by the band structure of the optical lattice.

We performed numerical calculations that trace the evolution of a noninteracting gas, produced initially at zero momentum, within the lattice band structure.  The numerical results shown in Fig.\ \ref{fig:TimeDep} are for $\theta=0.44$ radians and a lattice depth of 87 kHz with no free parameters.  The calculation matches well with the observed time dependence of the diffraction asymmetry.

To provide an intuitive description of the coherent dynamics in $\mathcal{A}$ that we both observe and calculate, we consider the effect of a small inversion-symmetry-breaking perturbation to the band structure of an inversion-symmetric lattice potential.  The unperturbed Hamiltonian $\mathcal{H}_0$, which includes the kinetic energy and the inversion-symmetric lattice potential $V_s(\mathbf{r})$, has eigenstates $|\psi_{i, \pm}^{(0)}\rangle$ that are either even (labeled by $+$) or odd (labeled by $-$) under the action of the spatial inversion.  The perturbation $\mathcal{H}_a$ results from the small antisymmetric lattice potential $V_a(\mathbf{r)}$ and mixes the even and odd eigenstates.  To first order in $
\mathcal{H}_a$, the zero quasimomentum eigenstates become

\begin{eqnarray}
\ket{\psi_{i,+}^{(1)}} &\approx& \ket{\psi_{i,+}^{(0)}} + \sum_j \alpha_{j,i}\ket{\psi_{j,-}^{(0)}} \\
\ket{\psi_{j,-}^{(1)}} &\approx& \ket{\psi_{j,-}^{(0)}} + \sum_i -\alpha_{j,i}^*\ket{\psi_{i,+}^{(0)}}
\end{eqnarray}
where $\alpha_{j,i}=\frac{\bra{\psi_{j,-}^{(0)}}\mathcal{H}_a \ket{\psi_{i,+}^{(0)}}}{E_{j,-}^{(0)} - E_{i,+}^{(0)}}$.

The initial state is a zero-momentum condensate that can be written in the basis of inversion-even eigenstates as $|\psi(0)\rangle = \sum_i c_i |\psi_{i, +}^{(0)}\rangle$. During the lattice pulse time $\tau$ this initial state evolves in time as
\begin{align}
\label{eq:TimeEv}
\ket{\psi(t)} = \sum_i c_i e^{-i \omega_{i,\small{+}} t} \left(  \ket{\psi_{i,+}^{(0)}} +  \sum_{j}\alpha_{j,i}\ket{\psi_{j,-}^{(0)}} \right) \\ \nonumber
+\sum_{j,k}-\alpha_{j,k}c_k e^{-i \omega_{j,\text{\small-}} t} \ket{\psi_{j,-}^{(0)}}
\end{align}
where $\omega_{i,\small{+}} = E_{i,+}/\hbar$ and $\omega_{j,\text{\small-}} = E_{j,-}/\hbar$.

The first term of Eq.\ $\left( \ref{eq:TimeEv} \right)$ represents the incorporation of inversion antisymmetry into the initially even eigenstates, and the second term represents fully antisymmetric states for which the perturbation introduces population at zero momentum.  Figure \ref{fig:ExpFig}(c) illustrates each of these effects on two states at zero quasimomentum that are heavily influenced by the perturbation $\mathcal{H}_a$: the initially symmetric ground state and antisymmetric first excited state.

The numerator of the inversion-asymmetry parameter $\mathcal{A}$ is the expectation value of an inversion-odd operator $M$ that is diagonal in the basis of reciprocal lattice momenta, with matrix element $\pm 1$ for the wave vectors $\pm \mathbf{G}_i$.  Using the first-order expression above for $|\psi(t)\rangle$, we obtain $\langle M \rangle = M_1(t) + M_2(t)$ with
\begin{eqnarray}
\label{eq:M}
M_1(t) & = & \sum_{i,j,k} \left( c^*_i c_k e^{-i \left(\omega_{k,\small{+}} - \omega_{i,\small{+}}\right)t} \alpha^*_{j,i} M_{j,k} + \text{c.c.} \right) \\
M_2(t) & = & \sum_{i,j,k} \left( c^*_k c_i e^{-i \left(\omega_{j,\small{-}} - \omega_{k,{\small{+}}}\right)t}  \left(-\alpha_{j,i}\right) M^*_{j,k} + \text{c.c.} \right) \nonumber
\end{eqnarray}
and $M_{j,i} = \langle \psi_{j,-}^{(0)} | M | \psi_{i,+}^{(0)}\rangle$.

These expressions identify two generic scenarios that lead to a large momentum-space asymmetry.  The first results in oscillations described by both $M_1(t)$ and $M_2(t)$ and involves a trio of eigenstates of the unperturbed Hamiltonian $\mathcal{H}_0$ at zero quasimomentum: two inversion-symmetric, $\ket{\psi_{i,+}^{(0)}}$ and $\ket{\psi_{k,+}^{(0)}}$, and one inversion-antisymmetric, $\ket{\psi_{j,-}^{(0)}}$.   These states can be identified by three key features.  First, the symmetric states have significant population at zero momentum so as to overlap with the stationary condensate, giving large $c_i$ and $c_k$.  Second, the inversion-antisymmetric state is close in energy to one of the inversion-symmetric states, say $\ket{\psi_{i,+}^{(0)}}$, so that $\alpha_{j,i}$ is large and they are strongly mixed by the perturbation $\mathcal{H}_a$.   Finally, the inversion-antisymmetric state and at least one of the inversion-symmetric states, say $|\psi_{k,+}^{(0)}\rangle$, have large population in the first-order diffraction momenta, so that $M_{j,k}$ is large.  When these criteria are satisfied, we expect modulations of equal strength in $M$ (and thus in $\mathcal{A}$) at frequencies $\omega_{k,\small{+}} - \omega_{i,\small{+}}$ and $\omega_{j,\small{-}} - \omega_{k,\small{+}}$.  The second scenario is described by $M_2(t)$ when $k=i$ and involves just two states -- $\ket{\psi_{i,+}^{(0)}}$ and $\ket{\psi_{j,-}^{(0)}}$.  These states are again characterized by large $c_i$ and $\alpha_{j,i}$, and must both have large population in the first-order diffraction momenta so that $M_{j,i}$ is large.  This scenario results in a modulation of $\mathcal{A}$ at  frequency $\omega_{j,\small{-}} - \omega_{i,\small{+}}$.   

In Fig.\ \ref{fig:ExpFig}(d) we show that just one trio of states in this perturbation picture explains most of the dynamical variation in $\mathcal{A}$.  Figure \ref{fig:ExpFig}(c) shows that the state $\ket{\psi_{1,+}^{(0)}}$ has large population in the zero and first-order diffracted momenta, that $\ket{\psi_{2,-}^{(0)}}$ has large population in the first-order momenta, and that these states are heavily mixed by the perturbation, i.e.,\ that $\alpha_{2,1}$ is large.  As a result, these two states are dominant contributors to oscillation in $\mathcal{A}$ as in the second scenario described, and also couple with a third state $\ket{\psi_{k,+}^{(0)}}$ as in the first scenario.  In Fig.\ \ref{fig:ExpFig}(d) we isolate the symmetric excited state with the largest population in the zero and first-order diffracted momenta, $|\psi_{31,+}^{(0)}\rangle$.  The energy of these three states define three frequencies that dominate the time evolution of $\mathcal{A}$.  The large momentum-space asymmetry is observed when the Kapitza-Dirac pulse time is tuned so that these temporal oscillations interfere constructively.  We note that there are several other symmetric excited states besides $|\psi_{31,+}^{(0)}\rangle$ that also play the role of $|\psi_{k,+}^{(0)}\rangle$ in the scenario we have outlined, and provide somewhat smaller contributions to the overall dynamics.

The observations and theoretical descriptions offered in this work illustrate how matter-wave diffraction can be made highly sensitive to, and strongly manipulated by, fine features of an optical lattice.  Our work also suggests an explanation for the momentum-space asymmetry observed in the diffraction of a Bose-Einstein condensate of two spin states of $^{87}$Rb and released from a spin-dependent optical lattice reported in Ref.\ \cite{solt12twisted} (see also Ref.\ \cite{jurg15twisted}).   The asymmetry was interpreted as evidence of a ground-state superfluid that forms with a spatially dependent phase in the superfluid order parameter.  A later theoretical study \cite{chou13twisted} found no evidence for such a ``twisted superfluid'' state, which is consistent with naive expectations given that the optical lattice and mean-field interaction potentials experienced by the atoms are both real valued.  

We suggest that the inversion-asymmetric diffraction patterns observed in the experiment \cite{solt12twisted} may have resulted from matter-wave diffraction from the inversion-asymmetric transient honeycomb lattice that repulsion from one atomic spin state creates for the second spin state.  Such a transient lattice potential would have an interaction-energy asymmetry between the A and B sites of the honeycomb lattice that is on the order of the superfluid chemical potential (around $h \times 1$ kHz).  This potential would persist for a time somewhat less than the recoil time (i.e.,\ around 100 $\mu$s).  The strength and duration of this asymmetric potential are comparable to those studied in the present work.  The interaction-driven diffraction of one matter wave off another can can be described equivalently as nonlinear coherent wave mixing induced by interatomic interactions \cite{pert10mixing}.  The observation in Ref.\ \cite{solt12twisted} that the sign of the asymmetry parameter $\mathcal{A}$  was consistent between experimental repetitions supports our view that the asymmetry resulted from deterministic matter-wave dynamics rather than by spontaneous symmetry breaking at a phase transition.  Moreover, in a recent experiment with the same system as in Ref.\ \cite{solt12twisted}, the diffraction was modified by eliminating one spin population from the lattice just before the atoms were released \cite{PhysRevA.93.033625}.  The consequent elimination of the asymmetry signal is consistent with our suggested explanation.

This work was supported by the NSF and the AFOSR through the MURI program.

\bibliographystyle{apsrev}

\bibliography{allrefs_x2}

\end{document}